\documentclass[letterpaper]{article} 
\usepackage{aaai2026}  
\usepackage{times}  
\usepackage{helvet}  
\usepackage{courier}  
\usepackage[hyphens]{url}  
\usepackage{graphicx} 
\usepackage{natbib}  
\usepackage{caption} 
\usepackage{algorithm}
\usepackage{algorithmic}

\usepackage{newfloat}
\usepackage{listings}
\DeclareCaptionStyle{ruled}{labelfont=normalfont,labelsep=colon,strut=off} 
\floatstyle{ruled}
\newfloat{listing}{tb}{lst}{}
\floatname{listing}{Listing}
\usepackage[utf8]{inputenc} 
\usepackage[T1]{fontenc}    
\usepackage{url}            
\usepackage{booktabs}       
\usepackage{amsfonts}      
\usepackage{nicefrac}       
\usepackage{microtype}      
\usepackage{graphicx}
\usepackage{multirow}
\usepackage{amsmath}
\usepackage{dirtytalk}
\usepackage{multirow}
\usepackage{enumitem}
\usepackage{lipsum,booktabs}
\usepackage{caption}
\usepackage{subcaption}
\usepackage{booktabs} 
\usepackage{tabularx}

\title{Automatic Funny Scene Extraction from Long-form Cinematic Videos}

\author{
    Sibendu Paul,
    Haotian Jiang,
    Caren Chen
}
\affiliations{
    Amazon Prime Video, Seattle, USA\\
    \{sibendu, haotij, carechen\}@amazon.com
}

\begin{document}

\maketitle

\begin{abstract}

Automatically extracting engaging and high-quality humorous scenes from cinematic titles is pivotal for creating captivating video previews and snackable content, boosting user engagement on streaming platforms. Long-form cinematic titles, with their extended duration and complex narratives, challenge scene localization, while humor's reliance on diverse modalities and its nuanced style add further complexity.
This paper introduces an end-to-end system for automatically identifying and ranking humorous scenes from long-form cinematic titles, featuring shot detection, multimodal scene localization, and humor tagging optimized for cinematic content.
Key innovations include a novel scene segmentation approach combining visual and textual cues, improved shot representations via guided triplet mining, and a multimodal humor tagging framework leveraging both audio and text. Our system achieves an 18.3\% AP improvement over state-of-the-art scene detection on the OVSD dataset and an F1 score of 0.834 for detecting humor in long text. Extensive evaluations across five cinematic titles demonstrate 87\% of clips extracted by our pipeline are intended to be funny, while 98\% of scenes are accurately localized.
With successful generalization to trailers, these results showcase the pipeline’s potential to enhance content creation workflows, improve user engagement, and streamline snackable content generation for diverse cinematic media formats.

\end{abstract}

\section{Introduction}
\label{sec:intro}
With the rapid growth of video-streaming platforms, such as Netflix and Prime Video, there is an increasing need to enhance user engagement through personalized and engaging content. With Netflix boasting 300 million subscribers (15.9\% YoY growth) and Prime Video over 200 million (10\% annual growth), these platforms increasingly rely on personalized video ads, and short-form video content to enhance user experience. One such avenue is the creation of short, humorous video clips that can capture user attention and improve discoverability. Humor, an essential element of human interaction, can be a powerful tool in promotional content, capturing attention~\cite{garner2005humor, wanzer2010explanation}, building rapport~\cite{stauffer99}, and fostering trust~\cite{vartabedian1993humor}.
However, the current process of extracting high-quality humorous scenes from full-length cinematic titles is predominantly manual, requiring significant time and effort. To address this, we propose an end-to-end automated pipeline for funny scene extraction that leverages multimodal data and deep learning to reduce costs and expand coverage across movies, web series, and trailers. The pipeline analyzes shots~\footnote{A shot is a sequence of frames captured continuously by the same camera~\cite{sklar1993film}} from any cinematic title, groups them into semantically cohesive scenes, and then identifies humorous moments for efficient extraction of engaging, contextually relevant content — enhancing promotional content creation and supporting the marketing goals of streaming platforms.


We address three key challenges in extracting humorous scenes from long‑form cinematic content: (1) robust scene detection over complex temporal structures, (2) accurate humor identification within extended scenes, and (3) filtering of improper humor with reliable ranking of genuine humorous segments. Unlike short‑range video analysis tasks such as shot detection or action recognition, scene localization in long‑form videos requires reasoning over both short‑ and long‑range dependencies, yet suffers from scarce large‑scale annotations due to the high cost of labeling boundaries. While self‑supervised learning has advanced for images and short clips, its extension to long‑form content is hindered by simplistic data curation and limited multi-modal signal extraction.

To address these challenges, we propose a multi‑modal scene‑segmentation framework that integrates visual and text signals, operating even without audio or with only sparse key frames per each shot. For visual signal, we draw inspiration from face‑clustering~\cite{schroff2015facenet} to group similar shots into scenes, using triplet loss to capture contextual relationships. Unlike prior contrastive shot‑representation methods~\cite{ qian2021spatiotemporal, chen2021shot, mun2022bassl} that rely on artificial augmentations or heuristic pair generation and ignore scene structure, we introduce a \textit{guided triplet generation} strategy leveraging MovieNet‑SSeg~\cite{huang2020movienet} to improve intra‑scene similarity and inter‑scene discrimination. To address its sparse key‑frame coverage, we incorporate shot‑level captions as complementary text cues, enabling robust representation even without audio. This multi‑modal approach generalizes effectively to benchmarks such as OVSD~\cite{rotman2016robust}, demonstrating its suitability for real‑world long‑form scene localization.

Aside from the scene localization, humor detection in long‑form cinematic content is challenging due to its multi‑faceted nature, spanning words, gestures, prosodic cues, and contextual dependencies. Prior work largely focuses on single‑modal approaches for short texts or videos~\cite{yang2015humor, chen2017predicting}, limiting applicability to extended scenes. We propose a multi‑modal humor‑tagging pipeline that fuses audio cues (prosody, laughter) with textual understanding of context–punchline relationships, using a modified ColBERT architecture with tailored training and deployment for long text. To safeguard user experience, we integrate a guardrail audio‑tagging model to filter improper humor such as bullying or mockery. Finally, we introduce a heuristic‑based humor scoring mechanism to rank scenes, enabling streaming platforms to surface the most engaging promotional content.

Overall, our contributions can be summarized as follows:
\begin{itemize}[noitemsep,nolistsep]
    \item We introduce an end-to-end automatic pipeline for extracting funny scenes from long-form cinematic titles.

    \item We propose a multimodal scene segmentation method that encodes visual and textual cues from individual shots, leveraging triplet loss with MovieNet‑SSeg supervision to improve shot–scene associations.

    \item We design a novel shot encoder combining the X‑CLIP cross‑frame transformer with a DINO projection head, enabling efficient representation learning with minimal training (80K triplets over 25 epochs).

\item Our scene detection model achieves state‑of‑the‑art performance on MovieNet‑SSeg and improves AP on OVSD by 18.3\%, demonstrating strong generalization.

\item For humor identification, we develop a multi‑modal audio‑text model that achieves an F1 score of 0.834 and accuracy of 0.728, and integrate a guardrail audio‑tagging model with 100\% recall to filter improper humor.

\item We introduce a heuristic‑based humor scoring mechanism to rank scenes, with curator evaluations confirming 87\% humor detection accuracy and 98\% scene localization in full‑length titles.

\end{itemize}
The primary use case of this end-to-end pipeline is to extract humorous scenes for autoplay functionality, enabling customers to preview content with a funny clip when hovering over a title, enhancing engagement. In the future, these output clips could be leveraged for a \say{fast laugh} experience on mobile platforms. Versatile by design, the pipeline can adapt to various scene extraction needs, catering to diverse streaming audience preferences.

\section{Related Works}
\label{sec:related}
\paragraph{Scene Detection.} 
Prior Unsupervised methods~\cite{rui1998exploring} cluster shots using visual features, while supervised approaches~\cite{rao2020local} improve performance but rely on costly annotations and multiple predefined shot features. Self-supervised methods~\cite{dorkenwald2022scvrl, chen2021shot, mun2022bassl} learn shot representations without annotations but still require scene boundary labels for fine-tuning alongside carefully crafted pretext tasks during pretraining. We use available scene boundaries to mine triplets and apply triplet loss for shot contrastive pre-training.
  
\paragraph{Humor detection.} Earlier NLP research focused on humor prediction from text alone~\cite{annamoradnejad2024colbert, purandare2006humor, kiddon2011s}, while recent work~\cite{kayatani2021laughing, hasan2019ur, brown2021face} highlights the value of multimodal approaches. Although datasets like UR-FUNNY~\cite{hasan2019ur}, MHD~\cite{patro2021multimodal}, and MusTARD~\cite{castro-etal-2019-towards} enable such studies, they lack humor annotations for longer scenes. 

\paragraph{Video retrieval.} With the growth of streaming platforms, video retrieval has become vital for browsing and searching large archives. It includes video–text retrieval—matching videos to captions—explored by works like ClipBERT~\cite{lei2021less}, Frozen~\cite{bain2021frozen}, CLIP4Clip~\cite{luo2022clip4clip}, DRL~\cite{wang2024disentangled}, and JSFusion~\cite{yu2018joint}, and similarity-based retrieval, such as copy detection~\cite{jiang2014vcdb} and near-duplicate retrieval~\cite{song2011multiple, wu2007practical}. Our focus is on extracting video scenes that preserve contextual meaning, particularly humor events — a task demanding nuanced understanding and intricate contextual reasoning, far more challenging than retrieving clips by simple tags like celebrity, place, or action.

\section{Methodology}
\label{sec:method}
\begin{figure}
    \includegraphics[width=0.46\textwidth]{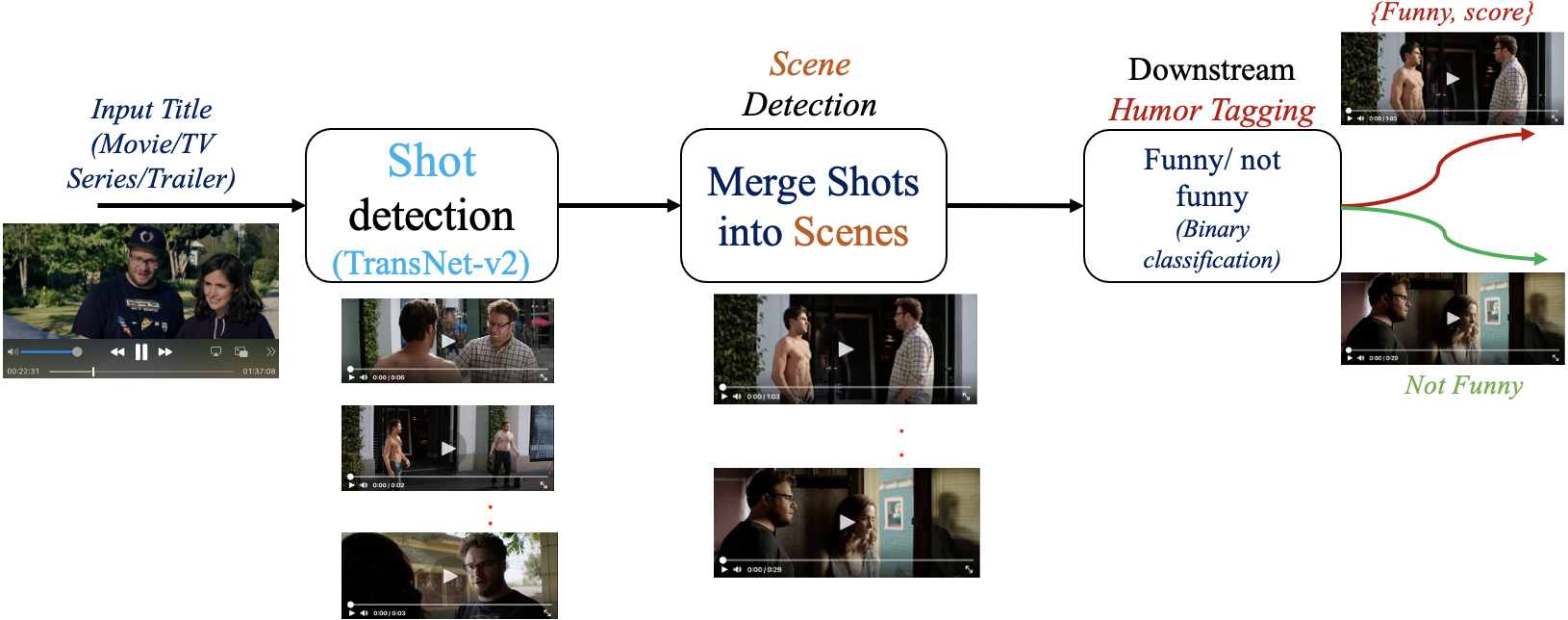}
\caption{\emph{Funny scene extraction pipeline overview.} The pipeline consists of three main blocks: shot detection, shot representation extraction, and scene merging, followed by downstream binary tagging and humor scoring.}
   \label{fig:pipeline} 
\end{figure}

Our proposed pipeline (shown in figure~\ref{fig:pipeline}) processes various titles e.g., movies, tv episodes etc, extracting funny scenes of variable length with assigned scores. 
Comprising three primary blocks, our pipeline first extracts shots, then merges semantically similar shots into scenes, and finally conducts binary humor classification and assign scores.


To extract frames captured from the same camera (i.e., shots), we utilize the state-of-the-art pretrained TransNetV2 network~\cite{souvcek2020transnet} as our shot-detector.

\subsection{Scene Detection}
\label{subsec:scenedetmethod}
A scene is defined as a sequence of shots depicting a semantically cohesive story segment. Unlike shot detection, which relies on visual cues, scene segmentation is more challenging due to the complex temporal relationships between shots~\cite{chen2021shot}. We propose a multimodal scene segmentation method that extracts and fuses visual and textual cues from each shot using separate encoders. The visual encoder is trained with self-supervised methods on task-specific data and pre-trained with Triplet Loss, while BLIP-2~\cite{li2023blip} serves as the textual feature extractor. Combined shot features are refined through supervised fine-tuning with neighboring context to accurately detect scene boundaries.

\subsubsection{Shot Contrastive pre-training}
We employ contrastive learning to obtain shot representations that capture local scene structures for scene boundary detection. Unlike prior methods, we treat shot-level representation learning as a multi-class classification pretext task, where each scene is a class and its shots are samples per class. Using Triplet Loss — a proven success in image-based face recognition~\cite{schroff2015facenet}, we learn distinctive shot embeddings suited for scenarios where the number of classes ($N_{Class}$) significantly outnumbers the number of samples for each class ($N_{SamplesPerClass}$). This parallels the face–person relationship in images with the shot–scene association in videos.


\paragraph{Triplet Mining.} Triplet Loss, though less greedy than other contrastive losses, depends heavily on selecting suitable anchor (s$_{i}^a$), positive (s$_{i}^p$) and negative (s$_{i}^n$)~\cite{tripletMining} samples. To generate triplets, we adopt an offline approach, leveraging ground-truth scene boundaries from MovieNet-SSeg train set~\cite{huang2020movienet}: shots within the same scene form positives, shots from different scenes form negatives, completing the triplet. To generate the hard triplets, we sample negative shots from the neighboring scenes within a $\pm$3-scene window. From 190 movies, we randomly sample 420 triplets per movie, yielding $\sim$ 80K triplets to pre-train our feature extractor. During pre-training, our goal is to make a shot (s$_{i}^a$) from a specific scene is closer to all other shots (s$_{i}^p$) from that scene than to any shot (s$_{i}^n$) from any other scene in the feature space, as in Equation~\ref{eq:triplet}, where $\alpha$ denotes the margin enforced between positive and negative pairs. We use a fixed $\alpha$ value of 1 throughout our training.


\begin{equation}   
    \lVert f(s_{i}^a) - f(s_{i}^p) \rVert_2^{2} + \alpha < \lVert f(s_{i}^a) - f(s_{i}^n) \rVert_2^{2},\ \  \forall (s_{i}^a, s_{i}^p, s_{i}^n) \in T
\label{eq:triplet}
\end{equation}

We generate triplets from ground-truth labels for added supervision, rather than using pseudo labels, leaving online hard triplet mining and progressive sampling for future work. 

\begin{figure*}
  \centering
       \includegraphics[width=0.96\textwidth]{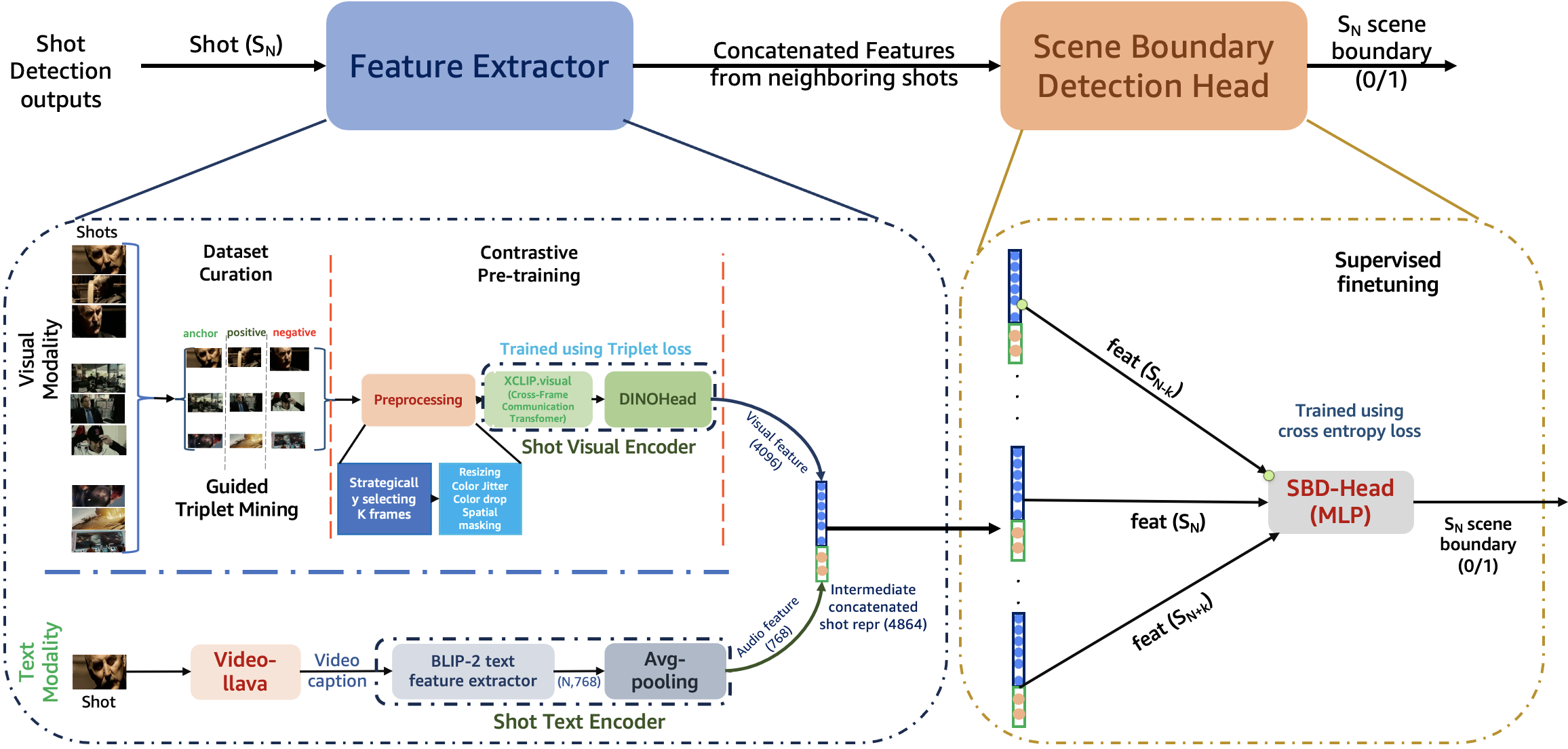}
       \caption{\emph{Scene detection module overview}: Triplets generated from MovieNet-SSeg ground-truth boundaries; spatial and temporal augmentations applied during contrastive pretraining with triplet loss; for each video, we extract shot-level video caption and use shot text-encoder to extract shot-level text features; fine-tuning aggregates neighboring shot features and trains MLP layers via supervised learning.}
        \label{fig:scene_det}
\end{figure*}

\paragraph{Shot Visual Encoder network.} Our shot encoder learns shot-level representations from visual modalities using the XCLIP visual encoder ~\cite{ni2022expanding} that leverages cross-frame attention to capture long-range temporal dependencies. To refine the transformer-extracted video feature representation, we append the DINO projection head~\cite{caron2021emerging} — a 3-layer MLP with GELU activations, L2-normalization, and a weight-normalized fully connected layer with 4096 dimensions. Hence, our shot encoder combines XCLIP’s cross-frame communication transformer with the DINO projection head, producing a 4096-dimensional feature per shot (Figure~\ref{fig:scene_det}).

\paragraph{Data Augmentation.} 
In video processing, shots require both temporal and spatial augmentations. For temporal augmentation, we use sparse sampling~\cite{wang2016temporal} to reduce redundancy. For spatial augmentation, we follow~\cite{dave2022tclr} with color jitter (contrast, hue, brightness, saturation), horizontal flipping, and spatial masking inspired by VideoMAE~\cite{tong2022videomae} to enhance shot encoder training. The impact of these augmentations on contrastive pre-training is examined in the \textit{Experiments} section.

\paragraph{Exploiting Text Modality.} For the large-scale scene detection dataset, where only a few key frames per shot are available, to explore the text-modality during the training, we use captions generated by vision-language model, video-llava~\cite{lin2023video} as the text input for each shot. These are encoded using a pre-trained BLIP-2~\cite{li2023blip} text extractor with mean pooling, yielding 768-dimensional text features. Each shot is then represented by concatenating the 4096-dimensional visual features with the 768-dimensional text features, forming a 4864-dimensional embedding.

\subsubsection{Supervised finetuning}

In long-range video analysis, determining scene boundaries requires considering neighboring shots. We use a window (W) covering N preceding and N succeeding shots, along with the center shot. Visual features from a triplet-loss trained shot encoder and textual features from BLIP-2 are concatenated per shot, then flattened into a ($2\times N + 1) \times 4864$ vector, fed to four-layer MLP SBD head ($[2\times N + 1]$$\times$4864-8192-4096-1024-2), with dropout applied after each of the first three FC layers. The window slides shot-by-shot, marking the center shot as a boundary if its score exceeds 0.5. During finetuning, the shot encoder is frozen and only the MLP is trained with cross-entropy loss.

\vspace{-0.1in}
\subsection{Downstream Scene Tagging}
\label{subsec:humortagging}

\begin{figure}
\includegraphics[width=0.47\textwidth]{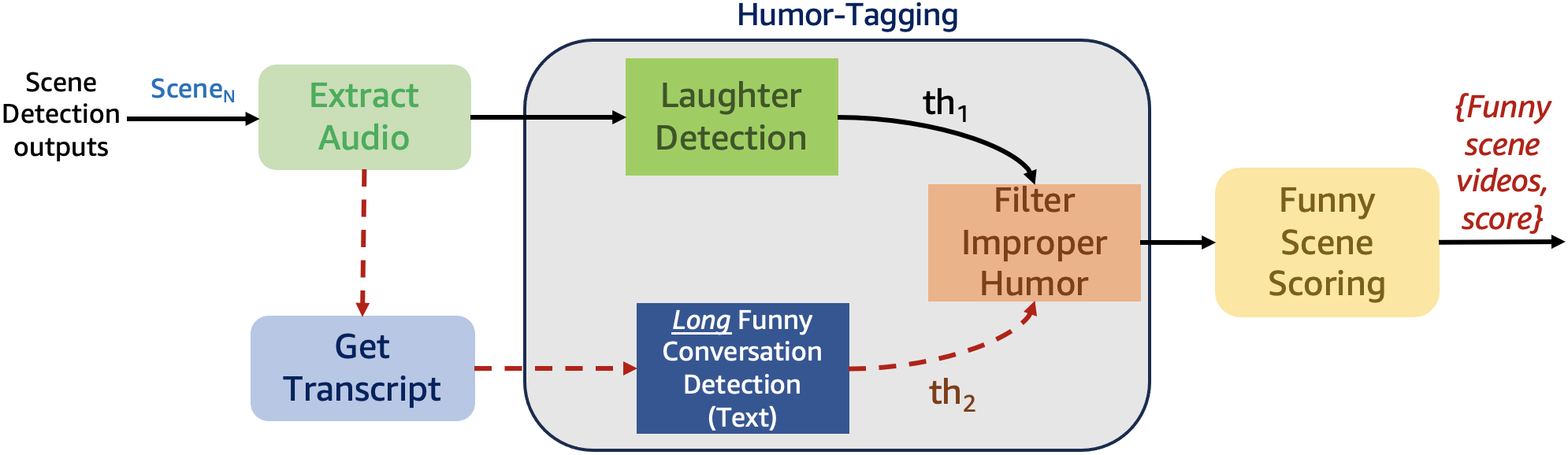}
    \caption{\emph{Multi-modal humor tagging pipeline overview}. \textbf{(1)} Extract audio and detect laughter, \textbf{(2)} transcribe audio and analyze funny conversations, \textbf{(3)} filter improper humor, then score and and output the ranked funny scene videos.}
    \label{fig:humor_tag}
    \vspace{-0.2in}
\end{figure}

Next, we label each scene to determine its humor content, focusing on eliciting positive emotions like amusement and laughter~\cite{funnydef}.
Our humor tagging approach utilizes audio and text analysis, detecting laughter in audio and using AWS-Transcribe with a custom-trained colBERT model to identify humor from verbal content. Audio from selected scenes after both analysis is further checked by an audio-tagging model to exclude scenes with improper humor. The filtered results are merged, and each humorous scene is scored to produce a ranked list of top humorous moments, as shown in Figure~\ref{fig:humor_tag}.

Recognizing the significance of laughter in identifying humor, we extract audio and detect it using a ResNet18-based model trained on the Switchboard dataset~\cite{godfrey1997switchboard}. We finetuned the model to accurately detect laughter from both individuals and background audiences.

\subsubsection{Humor detection from verbal modality.} 
According to Cambridge Dictionary, humor entails \say{the ability to be amused by something seen, heard, or thought about, sometimes causing you to smile or laugh, or the quality in something that causes such amusement}. Not all instances of amusement result in laughter, yet they may still be perceived as funny. This insight motivates us to extract humor from the verbal or text modality. Linguistically, humor often follows a setup, multi-stage storytelling, and a punchline that subverts expectations~\cite{eysenck1942appreciation, suls1972two}. Following this, the ColBERT model~\cite{annamoradnejad2024colbert} encodes individual sentences and the full text, concatenates them, and processes them through MLP layers to capture relationships among sentences and also scrutinize word-level connections throughout the entire text to assess the global congruity.

\paragraph{Model architecture and training.} The state-of-the-art ColBERT processes the first 5 input sentences along with the entire text, which works for short texts but is less effective for longer scenes (30s–2min) from cinematic titles. To address this, we adapt ColBERT by training on the UR-FUNNY dataset~\cite{hasan2019ur} which has the longest average scene duration (33.8s) among humor datasets. Our version processes 10 sentences plus the full text to capture broader context–punchline relationships, using effective sampling to improve inter-sentence relationship learning.

\paragraph{Training Scenario.} 
Inspired by the inherent structure of humor, which typically consists of a setup and punchline, we developed a 10-sentence sampling approach for training. Our semi-deterministic method handles longer texts by preserving the first two sentences for context and the last two for the punchline, while randomly selecting six sentences from the middle with appropriate spacing. This structure-dependent sampling strategy demonstrated superior generalization performance compared to other tested approaches.

\paragraph{Testing Scenario.} 
During the testing phase, we maintain determinism by avoiding random sampling and instead segment longer texts into multiple subtexts of 10 sentences each. These subtexts are processed individually through our custom-trained ColBERT model (trained on UR-FUNNY dataset), and their softmax scores are averaged to produce the final classification score. We employ an empirically determined softmax threshold of \textit{0.56} for humor detection to minimize false positives.


\subsubsection{Filtering out improper humor.}
Merely detecting laughter or humorous context alone is insufficient to ascertain the humorous quality of a scene, as it might include instances of improper humor such as bullying, mocking, or taunting. To tackle this, we employ a state-of-the-art pretrained audio-tagging CNN model~\cite{kong2020panns} trained on the AudioSet dataset. Using this model, we filter out inappropriate content by detecting scenes containing distress signals (crying, screaming) or negative emotional cues, thereby ensuring only appropriate humorous content reaches viewers~\footnote{Our current guardrail focuses on audio-based moderation, which effectively captures inappropriate humor violations such as violent cues. However, we acknowledge that a comprehensive safety system would benefit from multimodal guardrails that incorporate visual and textual modalities.}.

\subsubsection{Humor Score Generation.} 
We developed a heuristic scoring system for funny scenes using curator feedback as ground truth. It combines four normalized features—average laughter ($f_1$), laughter duration above threshold ($f_2$), ColBERT softmax score ($f_3$), and scene length ($f_4$)—weighted via grid search (Equation~\ref{eq:humor}). This enables effective ranking of top-k scenes against curator annotations for prioritizing humorous content for streaming media customers.

\begin{equation}   
    \footnotesize humor_{score} = (w_1 \times f_1) + (w_2 \times f_2) + (w_3 \times f_3) + (w_4 \times e^{-f_4/t_{c}})
\label{eq:humor}
\end{equation}

We acknowledge that humor is subjective and culturally variable, and our heuristic scoring may not fully capture styles like sarcasm, wordplay, or region-specific comedic cues. In future work, we aim to develop learned ranking models trained on more diverse, demographically varied annotations to improve cross-cultural generalization.

\section{Experiments}
\label{sec:experiments}



\subsection{Evaluation Metrics}

We evaluate each subtask independently using standard metrics: F1 for shot detection, AP and F1 for scene localization, and binary accuracy with F1 for our custom-trained colBERT model. 
Due to limited end-to-end evaluation data, we conduct a user study with professional curators, reporting both binary accuracy and qualitative insights.


\subsection{Experimental Results}
\label{subsec:expr_res}
Our pipeline comprises three sequential blocks (Figure~\ref{fig:pipeline}), with the optimal model selected for each stage and earlier stages fixed to their best-performing versions. Experimental results report the performance of each stage individually. 


We observed that our shot detector, TransNetV2~\cite{souvcek2020transnet}, outperforms AUTOSHOT~\cite{zhuautoshot} by 1.2\% on average F1 metric, even after a finetuned threshold on the BBC Planet Earth video dataset~\cite{bbc}.

\subsubsection{Scene Detection}
\label{subsubsec:scenedet}

We here use Average Precision (AP)~\footnote{AP value condenses the precision-recall (PR) curve into a single, confidence-threshold-independent metric that reflects the average of all precision values, making it well-suited to our use case.} and F1 score~\cite{chen2021shot}.

\paragraph{Impact of guidance while triplet generation.} \label{para:guidance}
\begin{table}
\centering{
\begin{tabular}{|p{2.0cm}|p{0.7cm}|p{0.7cm}|p{0.7cm}|p{0.7cm}|}
    \hline
    Shot-Encoder Variants & V1 & V2 & V3 & \textbf{final} \\
    \hline
    \hline
    \textit{NMI}-overall & 0.606 & 0.604 & 0.611 & \textbf{0.632} \\
    \hline
\end{tabular}
\caption{Scene clustering quality evaluated using normalized mutual information (NMI) across various triplet mining strategies.}
\label{tab:guidance}
}
\vspace{-0.2in}
\end{table}

\begin{table}
\scriptsize
\centering
{
\begin{tabular}{|p{0.7cm}|p{0.6cm}|p{0.6cm}|p{0.7cm}|p{0.7cm}|p{0.7cm}|p{0.4cm}|p{0.4cm}|}
    \hline
    Version & \multicolumn{5}{c|}{Augmentations} & best AP & best F1\\
    \cline{2-6}
    No. & Resize & Color & Horizon & Random & Uniform & & \\
    & & Jitter & tal Flip & Masking & Masking & & \\
    \hline
    A1 & \checkmark & \checkmark & - &  - & - & 41.7 & 35.1 \\
    \hline
    \textbf{A2} &  \checkmark & \checkmark & \checkmark & \checkmark & - & \textbf{47.2} & \textbf{41.8} \\
    \hline
    A3 &  \checkmark & \checkmark & \checkmark & - & 8\%-15\%  & 38.1 & 16.8 \\
    \hline 
     A4 &  \checkmark & \checkmark & \checkmark & - & 16\%-25\%  &  34.3 & 15.2\\
    \hline
    
    \hline
\end{tabular}
}
\caption{Impact of different augmentations on Movienet-Sseg.}
\label{tab:augmentations}
\end{table}

To evaluate the effect of guidance in triplet mining on shot‑encoder training, we compare three random heuristic‑based strategies that disregard annotations. In V1, positive shots are chosen within a $\pm$3‑shot window and negatives are $\geq$10 shots away; in V2, the window is reduced to $\pm$2 shots and negatives are $\geq$15 shots away; in V3, inspired by MovieNet‑SSeg statistics, positives are within $\pm$1 shot and negatives are $\geq$30 shots away. Each variant is used to pre‑train the shot encoder, and clustering quality is measured using normalized mutual information (NMI) over 100 randomly sampled scenes from the MovieNet‑test split. As shown in Table~\ref{tab:guidance}, V3 yields fewer false positives and better clustering than V1 and V2, while our guided triplet generation - eliminating false‑positive triplets and achieves the highest NMI, highlighting the value of incorporating guidance in triplet mining for improved shot‑representation learning.

\paragraph{Impact of different augmentations.}
We apply various data augmentations during shot contrastive pretraining to enhance representation robustness. Table~\ref{tab:augmentations} compares four combinations (A1–A4), with A2—featuring resizing, color jitter, horizontal flipping, and random masking—achieving the best scene detection accuracy. The masking strategy, inspired by prior work, improves recall but reduces precision, with A2 striking the best balance to boost AP and F1.


\begin{table}
\small
    \centering
    \begin{tabular}{|c|c|}
    \hline
    Model & best AP \\
    \hline
    ShotCoL~\cite{chen2021shot} & 25.02 \\
    \hline
    BASSL~\cite{mun2022bassl} & 28.68 \\
    \hline
     \textbf{Ours (Visual)} & \textbf{33.94} \\
    \hline
\end{tabular}
\caption{Comparison between our method and other shot-level pre-training baselines on OVSD, utilizing visual signals (V) only.}
\label{tab:ovsd}
\end{table}

\paragraph{Generalization performance.}
MovieNet-SSeg~\cite{huang2020movienet} provides 42K scenes from 318 movies with 3 key frames per shot, while OVSD~\cite{rotman2016robust} offers continuous video input, aligning closely with our (Prime Video) deployment scenario. We extract 8 uniformly spaced frames per shot to reduce temporal redundancy. Table~\ref{tab:ovsd} shows that our model, trained on MovieNet-SSeg using only visual features, generalizes well to OVSD without finetuning.

\paragraph{Impact of incorporating Text Modality.}
Incorporating text-based features (768) alongside larger visual features (4096) significantly improves scene detection. As shown in Table~\ref{tab:impact-input-modality}, adding text boosts AP by 9.1\% and F1 by 11\% on the MovieNet-SSeg test set, enabling the model to leverage unique shot-level cues for better scene localization.

\begin{table}[h!]
\centering
{
\begin{tabular}{|c|c|c|}
    \hline
    Input-modality & best AP & best F1\\
    \hline
    Visual (V) & 47.2 & 41.8 \\
    \hline
     Visual (V) + Text (T) & 51.5 & 46.4 \\
    \hline
\end{tabular}
\caption{Impact of input modalities on the scene detection performance on Movienet-Sseg test set.}
\label{tab:impact-input-modality}
}
\vspace{-0.1in}
\end{table}

\subsubsection{Downstream Humor Tagging}
\label{subsubsec:humortag}

After optimizing the first two stages of the pipeline depicted in Figure~\ref{fig:pipeline}, we evaluate the performance of the custom-trained colBERT model for humor tagging based on text modality. Due to space constraints, the individual impact of the audio modality is omitted here.

\paragraph{\textbf{Using Text Modality. }}
We evaluated four model variants based on input sentence count (5 or 10) and embeddings (bert-base-uncased or roberta-base). The colBERT model trained on the UR-FUNNY dataset~\cite{hasan2019ur} using 10 input sentences and bert-base embeddings achieved the best generalization performance on the MHD dataset~\cite{patro2021multimodal}, with an F1 score of \emph{0.834} and an accuracy of \emph{0.728}. This outperforms other variants, the state-of-the-art finetuned transformer~\cite{weller2019humor}, and even multimodal fusion models like FunnyNet~\cite{liu2022funnynet}, despite using only text modality.

\paragraph{\textbf{Evaluation of Humor Score Generation. }}

To rate the humor in output scenes for viewer presentation, we propose a heuristic-based mechanism relying on four high-level signals: (1) average laughter score ($f_1$), (2) percentage of time laughter score exceeds a threshold ($f_2$), (3) custom-trained ColBERT model softmax score($f_3$), and (4) scene duration ($f_4$)~\footnote{Shorter funny scenes are preferred by professional curators.}. The weights [$w_1$, $w_2$, $w_3$, $w_4$] (used in equation~\ref{eq:humor}) are optimized using grid search across three regression models: linear, logistic, and decision-tree regression. 

For evaluation, we propose a custom intersection-over-union (IoU)-based metric combining coarse-grained and fine-grained estimations. The coarse-grained metric, $topIOU_N$, measures overlap between the top-N humorous scenes predicted by our pipeline and those identified by curators~\footnote{Output funny scenes scored and ranked by professional curators serve as the ground truth i.e., $\{gt\}$.} (Equation~\ref{eq:topIOU}). The fine-grained metric, $topIOU\_Align_N$, evaluates alignment between the top-N predicted funny scenes and the curators’ top-N ranked scenes (Equation~\ref{eq:topIOUAlign}). Both metrics are normalized for three N values (3, 5, 10) and aggregated into an overall evaluation metric, $eval_{metric}$, as shown in Equation~\ref{eq:humoreval}.

\begin{equation}   
    eval_{metric} = \sum_{N=\{3,5,10\}} (topIOU_N + topIOU\_Align_N) \\
    \label{eq:humoreval}
\end{equation}   
\begin{align} 
    topIOU_N &= |\{gt[:]\} \cap \{predicted[:N]\}|/N \label{eq:topIOU}\\
    topIOU\_Align_N &= |\{gt[:N]\} \cap \{predicted[:N]\}|/N \label{eq:topIOUAlign}
\end{align}

Using 60\% of curator-annotated samples for weight estimation and testing on the remaining 40\%, we found logistic regression achieves the best $eval_{metric}$ score on the test set.


\begin{table}[h!]
\centering
\setlength{\tabcolsep}{4pt} 
\begin{tabular}{|c|c|p{2.6cm}|p{1cm}|c|}
    \hline
    Train Set & Test Set & Model & Acc & F1 \\
    \hline
    \multirow{5}{2.2em}{UR-FUNNY} & \multirow{5}{1.8em}{MHD} & Fine-tuned Transformer & 0.591 & 0.719 \\
    \cline{3-5}
    & & ColBERT-base-5 & 0.694 & 0.820 \\
    \cline{3-5}
    & & \textbf{ColBERT-base-10} & \textbf{0.728} & \textbf{0.834} \\
    \cline{3-5}
    & & ColRoBERTa-base-5 & 0.720 & 0.830 \\
    \cline{3-5}
    & & ColRoBERTa-base-10 & 0.690 & 0.812 \\
    \hline
    \hline
    MHD & MHD & MASM~\cite{patro2021multimodal} & 0.720 & 0.810 \\
    \hline
\end{tabular}
\caption{Our text-based humor tagging model performance.}
\label{tab:mhd}
\vspace{-0.1in}
\end{table}

\subsection{Professional curators' Evaluation} 
To comprehensively evaluate our funny scene generation pipeline, we tested it on five $\sim$ 2-hour films spanning comedy, drama, action, adventure, and animation, as well as on 11 trailers from diverse genres. The pipeline (Figure \ref{fig:pipeline}) produced $\sim$ 100 scenes from the films and 17 scenes (10–15s each) from the trailers, all reviewed by three professional curators. Curators assessed each clip for objective humor (clear audio-visual or textual cues such as in-context laughter, smiles, or on-screen jokes, excluding improper content like bullying or cultural stereotypes, and including subtle forms like sarcasm and witty remarks), subjective humor (whether the clip felt funny regardless of explicit cues or creator intent), and scene ending quality (a natural, complete conclusion without abrupt cuts or unresolved dialogue).

Results (Table \ref{tab:ops}) show 98\% scene localization accuracy for main content, with higher segmentation errors in trailers due to rapid transitions. Under objective guidelines, humor detection accuracy reached 87\% for films and 100\% for trailers; subjective evaluation yielded 74\% and 88\%, respectively. While trailers had more segmentation errors, their condensed, event-rich nature made humor easier to detect. Inter-annotator agreement, measured by pairwise Cohen’s $\kappa$, had a minimum value of 0.75,  indicating substantial agreement among curators in assessing the humorous nature of the clips. Our approach extracts humorous scenes independent of genre, and we currently use a simple heuristic that skips the final 20\% of a title to avoid spoilers. A more sophisticated spoiler-aware mechanism would further improve robustness across all genres.


\begin{table}[h!]
\small
\centering
{
\begin{tabular}{|c|c|c|}
    \hline
    & Movies & Trailers \\
    \hline
     Is the clip intended to be funny? & 87\%  & 100\% \\
    \hline
   Do curators find the clip funny? &  74\% & 88\%\\
    \hline
    Proper Scene Ending?  & 98\% & 53\%\\
    \hline
    
\end{tabular}
}

\caption{curators' evaluation on movies \& trailers.}
\vspace{-0.2in}
\label{tab:ops}
\end{table} 

\begin{figure}[t]
\centering

\subcaptionbox{Output funny scene based on funny conversation detection.\label{fig:example_1}}{
    \includegraphics[width=0.95\linewidth]{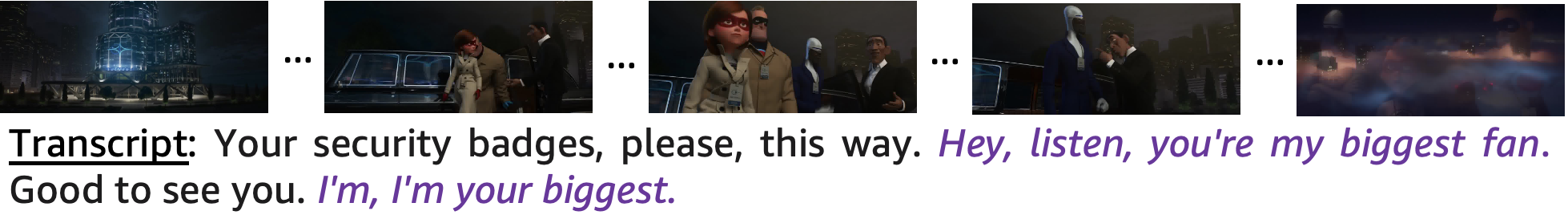}
}
\vfill
\subcaptionbox{Laughter and funny conversation detected with ``moaning, grunting'' audio labels (instance of bullying).\label{fig:improper_2}}{
    \includegraphics[width=0.95\linewidth]{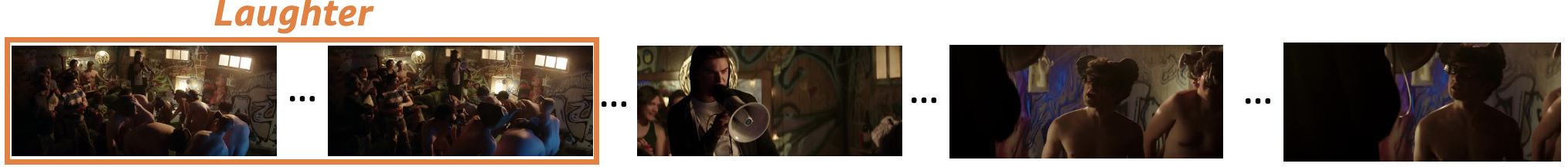}
}

\vfill
\subcaptionbox{Output funny scene based on laughter detection.\label{fig:example_3}}{
    \includegraphics[width=\linewidth]{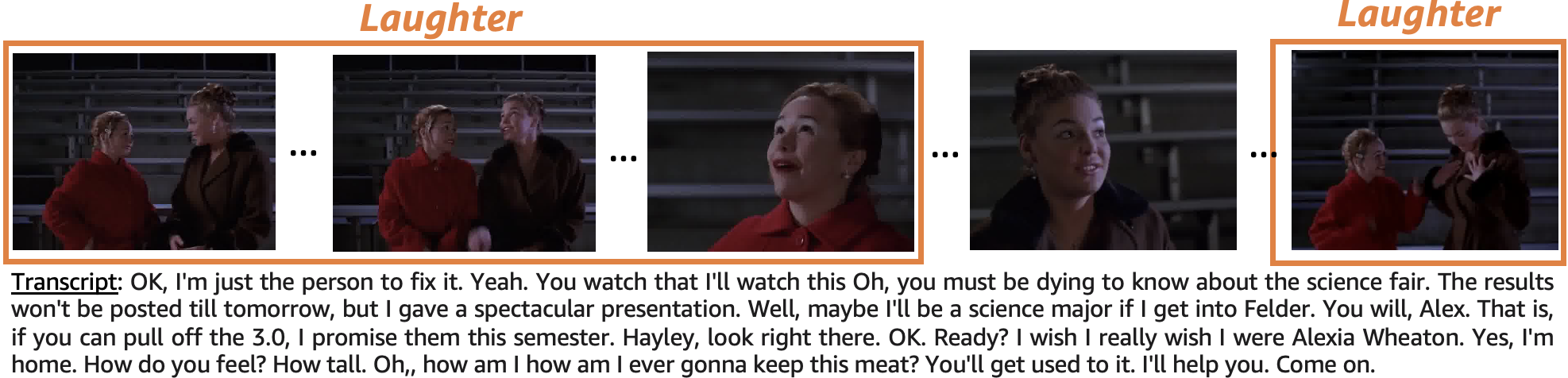}
}
\vfill
\subcaptionbox{Output funny scene based on both laughter and funny conversation detection.\label{fig:example_5}}{
    \includegraphics[width=\linewidth]{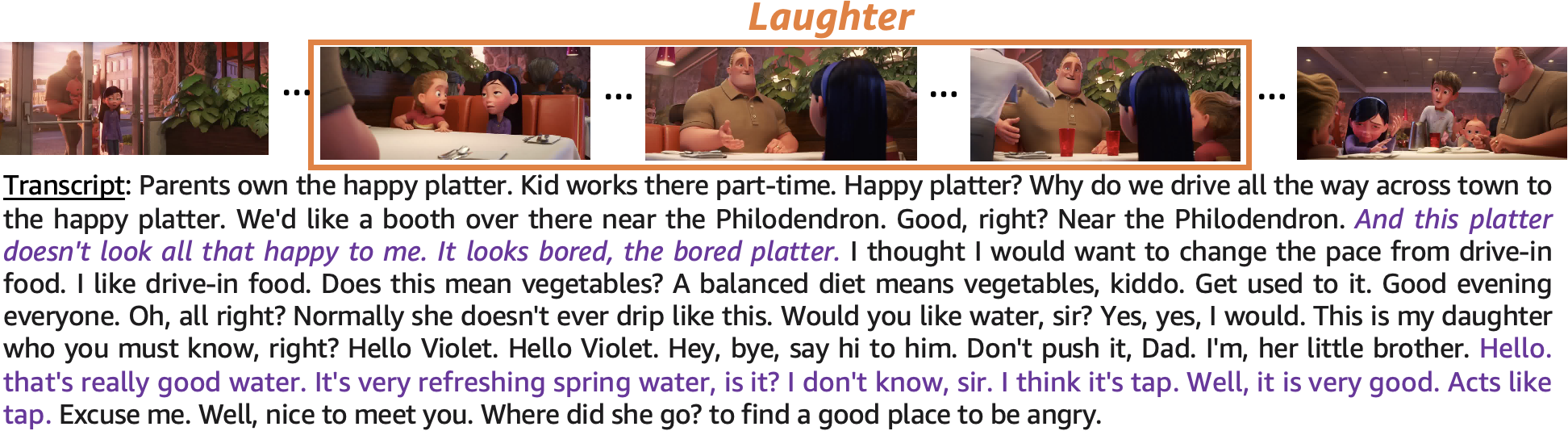}
}
\caption{Example extracted humorous scenes and instances of improper humor from various long-form cinematic titles. Observed laughter is marked with \textcolor{orange}{orange} boxes, while laugh-worthy dialogue or humorous context is highlighted in \textcolor{violet}{\textit{violet}}.}
\label{fig:funny_examples}
\vspace{-0.2in}
\end{figure}

\paragraph{\textbf{Qualitative Analysis.}}
Here, we present example funny scenes extracted by our proposed pipeline using multimodal cues. Figure~\ref{fig:example_1} and \ref{fig:example_5} show funny scenes from an action/adventure anime movie, detected using our multimodal pipeline. In Figure~\ref{fig:example_1}, humor arises from incongruity as a character contradicts himself—first calling a guest his fan, then claiming to be the guest’s biggest fan. In Figure~\ref{fig:example_5}, the pipeline captures both laughter and witty dialogue, including mocking a restaurant named “Happy Platter” as “Bored Platter” and joking about its water. Figure~\ref{fig:example_3} presents a comedy/family movie scene identified through repeated laughter cues. In contrast, Figure~\ref{fig:improper_2} highlights improper humor, where students mock another’s distress, detected via audio cues.


\section{Pathway to Deployment \& Future Work}
\label{sec:deploy}

The end-to-end automatic humor scene extraction pipeline is being deployed in phases, with TransNet-v2 shot detection and multi-modal scene detection modules already operational and supporting several use cases, including the generation of compelling, socially shareable personalized scenes designed to boost Prime Video user engagement. The final phase involves deploying the humor tagging pipeline, which requires meticulous testing and robust guardrails to ensure high precision in identifying appropriate humorous content, prioritizing accuracy over recall to maintain a positive viewing experience for Prime Video customers.


Our method performs strongly on long‑form content but struggles with rapid transitions in trailers; future work will explore adaptive window sizes and enhanced temporal modeling to address this. We will also investigate progressive triplet‑hardness scheduling to improve shot‑representation generalization. 

While our ColBERT humor detection model currently supports only English language content, we plan to expand into multilingual humor detection and conduct comprehensive user studies to minimize evaluation bias and enhance global accessibility.

\section{Discussion \& Future work}
Our proposed multimodal scene segmentation performs well for long-form cinematic titles but faces challenges in fast-paced trailers due to rapid transitions. To address this, we propose adaptive window sizes for neighboring shot representations or enhanced temporal modeling. The adaptive approach employs a dynamic moving window (W) with multi-scale sizes (1, 2, and 4), prioritizing immediate neighbors while accommodating varying scene lengths. This method improves boundary detection across diverse shot dynamics. 

Enhanced temporal modeling can be achieved through improved dataset curation during training. Our visual shot contrastive pre-training uses guided triplet generation with varying hardness levels. Selecting the hardest negatives early can cause model collapse due to bad local minima. To counter this, we propose gradually increasing triplet hardness by reducing the $\alpha$ value in equation~\ref{eq:triplet} during training~\cite{schroff2015facenet}. While this adds complexity, online triplet mining enhances scene representation learning by improving tolerance to intra-class variance and preventing embedding space over-clustering~\cite{xuan2020improved, sikaroudi2020offline}, leading to better generalization. Future work will explore these strategies to improve scene segmentation and enhance funny scene extraction performance.
 

%

Due to the lack of annotated datasets for end-to-end evaluation, we rely on assessments by professional curators. As humor is inherently subjective, this may introduce bias; however, we mitigate this through standardized evaluation protocols and consistent guidelines. Future work will include broader user studies to improve reliability. 

While laughter detection and audio-based improper humor filtering support non-English titles, our text-based funny conversation detection currently relies on English transcriptions. Since humor varies across languages, future work will focus on designing language-specific humor detection models to support global streaming catalogs.

\section{Conclusion}

In this work, 
we present an end-to-end system for automatically extracting humorous scenes from long-form cinematic content to generate engaging previews. By combining shot detection, multimodal scene localization using visual and caption-based representations, and multimodal humor tagging, the system significantly improves performance -- achieving an 18.3\% gain on the OVSD dataset and surpassing state-of-the-art humor detection baselines. Professional curator evaluations further validate the approach with 87\% accuracy on main content, while the modular design enables easy adaptation to diverse scene extraction and personalization tasks.
\bibliography{sample}

\end{document}